\documentclass[final,5p,times,twocolumn]{elsarticle}

\usepackage{amsfonts,amsmath,amssymb,bm}
\usepackage{graphicx} 
\usepackage{color}
\usepackage{textcomp}

\definecolor{prdblue}{rgb}{0.133,0.118,0.498}
\usepackage[colorlinks=true, pdfstartview=FitV, linkcolor=prdblue, citecolor= prdblue, urlcolor=prdblue]{hyperref}

\newcommand{\be}{\begin{equation}}
\newcommand{\bea}{\begin{eqnarray}}
\newcommand{\ee}{\end{equation}}
\newcommand{\eea}{\end{eqnarray}}

\def\spr{\!\cdot\!}

\def\noeq#1{(\ref{#1})}
\def\1eq#1{Eq.~(\ref{#1})}

\def\2eqs#1#2{Eqs.~(\ref{#1}) and~(\ref{#2})}
\def\3eqs#1#2#3{Eqs.~(\ref{#1}),~(\ref{#2}) and~(\ref{#3})}

\def\fig#1{Fig.~\ref{#1}}

\def\ie{{\it i.e.}, }

\def\n#1{({\it #1})}

\def\t{\lambda}

\def\gl{Q}
\def\Zgl{Z_\gl}

\def\ffT{T}

\def\ffGT{\Gamma_T}
\def\ffGTR{\Gamma_{T\!,\,R}}

\def\ffTs{T^{\mathrm{sym}}}
\def\ffSs{S^{\!\mathrm{sym}}}
\def\ffGTs{\Gamma_T^{\mathrm{sym}}}
\def\ffGSs{\Gamma_S^{\mathrm{sym}}}

\def\ffTas{T^{\mathrm{asym}}}
\def\ffGTas{\Gamma_T^{\mathrm{asym}}}

\biboptions{sort&compress}
\journal{Physics Letters B}

\begin{document}

\begin{frontmatter}

\title{On the zero crossing of the three-gluon vertex}

\author[UoC]{A. Athenodorou}
\author[ECT]{D. Binosi}
\author[LPT]{Ph.~Boucaud} 
\author[UPO]{F.~De Soto}
\author[UV]{J.~Papavassiliou}
\author[UH]{J.~Rodr\'{\i}guez-Quintero}
\author[GUF]{S.~Zafeiropoulos}

\address[UoC]{Department of Physics, University of Cyprus, POB 20537, 1678 Nicosia, Cyprus}
\address[ECT]{European Centre for Theoretical Studies in Nuclear Physics and Related Areas (ECT*) and Fondazione Bruno Kessler, Villa Tambosi, Strada delle Tabarelle 286, I-38050 Villazzano (TN), Italy.}
\address[LPT]{ Laboratoire de Physique Th\'eorique (UMR8627), CNRS, Univ. Paris-Sud, Universit\'e Paris-Saclay, 91405 Orsay, France}
\address[UPO]{Dpto. Sistemas F\'isicos, Qu\'imicos y Naturales, 
Univ. Pablo de Olavide, 41013 Sevilla, Spain}
\address[UV]{Department of Theoretical Physics and IFIC,
University of Valencia-CSIC, E-46100, Valencia, Spain.}
\address[UH]{Department of Integrated Sciences;  
University of Huelva, E-21071 Huelva, Spain.}
\address[GUF]{Institut f\"ur Theoretische Physik, Goethe-Universit\"at Frankfurt,
Max-von-Laue-Str.~1, 60438 Frankfurt am Main, Germany}

\date{\today}

\begin{abstract}
%
%
We report on new results on the infrared behaviour of the three-gluon vertex in quenched Quantum Chormodynamics, obtained from large-volume lattice simulations. The main focus of our study is the appearance of the  characteristic infrared feature known as `zero crossing', the origin of which is intimately connected with the nonperturbative masslessness of the Faddeev-Popov ghost. The appearance of this effect  is clearly visible in one of the two kinematic configurations analyzed, and its theoretical origin is discussed in the framework of  Schwinger-Dyson equations. The effective coupling in the momentum subtraction scheme that corresponds to the three-gluon vertex is constructed, revealing the vanishing of the effective interaction at the exact location of the zero crossing.  

\end{abstract}

\begin{keyword}
Lattice simulations\sep
three-gluon vertex\sep
zero crossing\sep
Schwinger-Dyson equations
\smallskip

\end{keyword}
\end{frontmatter}


\noindent\textbf{1.$\;$Introduction}. 

One notable aspect of the ongoing intense exploration of the 
 infrared (IR) sector of Quantum Chromodynamics (QCD) 
has been the detailed scrutiny 
of the fundamental Green's functions
of the theory using large-volume lattice simulations~\cite{Cucchieri:2006tf,Cucchieri:2008qm,Cucchieri:2007md,Cucchieri:2010xr,Bogolubsky:2009dc,Oliveira:2009eh, Ayala:2012pb,Duarte:2016iko}, together with a variety              
of continuum approaches~\cite{Aguilar:2008xm,Boucaud:2008ky,Fischer:2008uz,RodriguezQuintero:2010wy,
  Pennington:2011xs,Maris:2003vk,Aguilar:2004sw,Boucaud:2005ce,Fischer:2006ub,Kondo:2006ih,Binosi:2007pi,Binosi:2008qk, 
Boucaud:2007hy,Dudal:2007cw,Dudal:2008sp,Kondo:2011ab,Szczepaniak:2001rg,Szczepaniak:2003ve,Epple:2007ut,Szczepaniak:2010fe,Watson:2010cn,Watson:2011kv}. 
Even though off-shell Green's functions  
are not physical quantities, given their explicit dependence on the 
gauge-fixing parameter and the renormalization scheme, 
they encode valuable information on fundamental  nonperturbative 
phenomena such as confinement, chiral symmetry breaking, and dynamical mass generation,
and constitute the basic building blocks of symmetry-preserving formalisms 
that aim at a veracious description of hadron  phenomenology
\cite{Maris:2003vk,Chang:2009zb,Chang:2011vu,Qin:2011dd,Qin:2011xq,Bashir:2012fs,Eichmann:2012zz,Cloet:2013jya,Binosi:2014aea}.

The most important findings of the aforementioned studies 
are related with the two-point sector of the theory. Specifically, 
it is now firmly established that, in the Landau gauge, 
the gluon propagator, $\Delta(q^2)$, reaches a finite value in the deep IR, 
whilst the ghost propagator, $D(q^2)$ remains massless, but with an 
IR finite dressing function, $F(q^2)$ [note that $D(q^2) =F(q^2)/q^2$].
This characteristic behavior has
led to the critical reassessment of previously established 
theoretical viewpoints, and has sparked a systematic effort towards a `top-down' 
derivation of the ingredients that enter in the dynamical equations 
describing the properties of mesons~\cite{Binosi:2014aea}.

The aforementioned results  
may be explained in a rather natural way 
within the framework of the Schwinger-Dyson equations (SDEs),  
by invoking the concept of a dynamically generated gluon mass~\cite{Cornwall:1981zr}.  
The self-consistent picture that emerges may be succinctly summarized 
by saying that~\cite{Binosi:2009qm} \n{\bf a} the gluon acquires an effective mass through a 
subtle realization of the Schwinger mechanism, the implementation of which hinges on  the dynamical formation of longitudinally-coupled poles, and 
\n{\bf b} the ghost is transparent to this mechanism, and remains 
massless; however, its dressing function is protected by the 
gluon mass, that tames any possible IR divergence and 
enforces its finiteness at the origin. 

This profound difference in the IR behaviour 
between gluons and ghosts induces characteristic effects 
to other Green's functions [but also to  $\Delta(q^2)$],
essentially due to the  
inequivalence between loops containing `massive' gluons 
or massless ghosts~\cite{Aguilar:2013vaa}. 
Specifically, while the former are `protected' by the gluon mass, $m$,  
yielding IR finite contributions of the type $\log(q^2+m^2)$, the latter 
are `unprotected', yielding (potentially) IR divergent terms of the type $\log q^2$. 

In the case of $\Delta^{-1}(q^2)$, the ghost-loop contained in its 
self-energy generates a term $q^2\log q^2$, and therefore remains IR finite; 
however, the first derivative of $\Delta^{-1}(q^2)$ diverges at the origin, precisely 
as an unprotected logarithm.  

The corresponding 
effect on the three-gluon vertex is 
particularly striking. Specifically, 
in certain special kinematic limits, some of the vertex form factors
are dominated in the IR by  the corresponding 
ghost-loop diagram, the leading contribution of which, by virtue of 
the Slavnov-Taylor identity (STI), turns out to be proportional 
to the derivative of $\Delta^{-1}(q^2)$. Thus, the form factors  
reverse sign for sufficiently small
momenta, displaying finally a logarithmic divergence at the origin.     
The transition from positive values 
(at intermediate and large momenta) to a negative divergence at the origin is 
associated with the so-called {\it `zero crossing'}: at some finite energy 
scale, in the deep IR, the form factors in question vanish.
The weak nature of the divergence makes the effect difficult to observe 
in lattice simulations. Indeed, SU(2) studies~\cite{Cucchieri:2006tf,Cucchieri:2008qm} found the expected
pattern in  three space-time dimensions (where the IR divergence is linear in $q$), 
but were less conclusive in four dimensions.

In  this letter,  we present new  results for  the
three-gluon  vertex obtained from SU(3)  lattice simulations 
in  large four-dimensional volumes. We restrict our analysis to the 
tensorial structure corresponding to that of the three-gluon vertex, 
which is obtained as a particular projection of the full lattice three-point 
function, after  the amputation of the external gluon legs.
The results strongly support the appearance of a zero crossing
in the case of one of the two kinematic configurations considered
(`symmetric' configuration). On the other hand, in the case of the second 
kinematic choice (`asymmetric'configuration), 
the presence of a zero crossing cannot be clearly discerned. 
The theoretical origin of this special feature is reviewed within the 
framework of the SDEs, and the three-gluon running coupling,  
defined in the momentum subtraction (MOM) scheme, numerically extracted from the data. Finally, the limitations 
of the semiclassical approach in accounting for the observed behavior 
of the three-gluon vertex in the IR are briefly discussed. 

\smallskip

\noindent\textbf{2.$\;$Three-gluon vertex, renormalization, and effective charge}. 
The connected three-gluon  vertex is defined as the correlation function ($q+r+p=0$)
\begin{align}
	{\cal G}^{abc}_{\alpha\mu\nu}(q,r,p)=\langle{\gl^a_\alpha(q)}{\gl^b_\mu(r)}{\gl^c_\nu(p)}\rangle=f^{abc}{\cal G}_{\alpha\mu\nu}(q,r,p),
\end{align}
where the sub (super) indices represent Lorentz (color) indices and the average $\langle \cdot \rangle$ indicates functional integration over the gauge space. In terms of the usual 1-particle irreducible (1-PI) function, one has
\begin{align}
	{\cal G}_{\alpha\mu\nu}(q,r,p)&=g\Gamma_{\alpha'\mu'\nu'}(q,r,p)\Delta_{\alpha'\alpha}(q)\Delta_{\mu'\mu}(r)\Delta_{\nu'\nu}(p),
	\label{Cto1-PI}
\end{align}
with $g$ the strong coupling constant. In the Landau gauge, the transversality of the gluon propagator, {\it viz.}, 
\begin{align}
	\Delta^{ab}_{\mu\nu}\left(q\right)=\langle \gl^a_\mu(q) \gl^b_\nu(-q) \rangle=  \delta^{ab} \Delta(p^2) P_{\mu\nu}(q),  
\end{align}
where $P_{\mu\nu}(q)=\delta_{\mu \nu} - q_\mu q_\nu/q^2$, implies directly that $\cal G$ is totally transverse: $q\spr{\cal G}=r\spr{\cal G}=p\spr{\cal G}=0$. 

\begin{figure*}[!t]
\begin{center}
	\hspace{-.75cm}
	\includegraphics[width=0.46\linewidth]{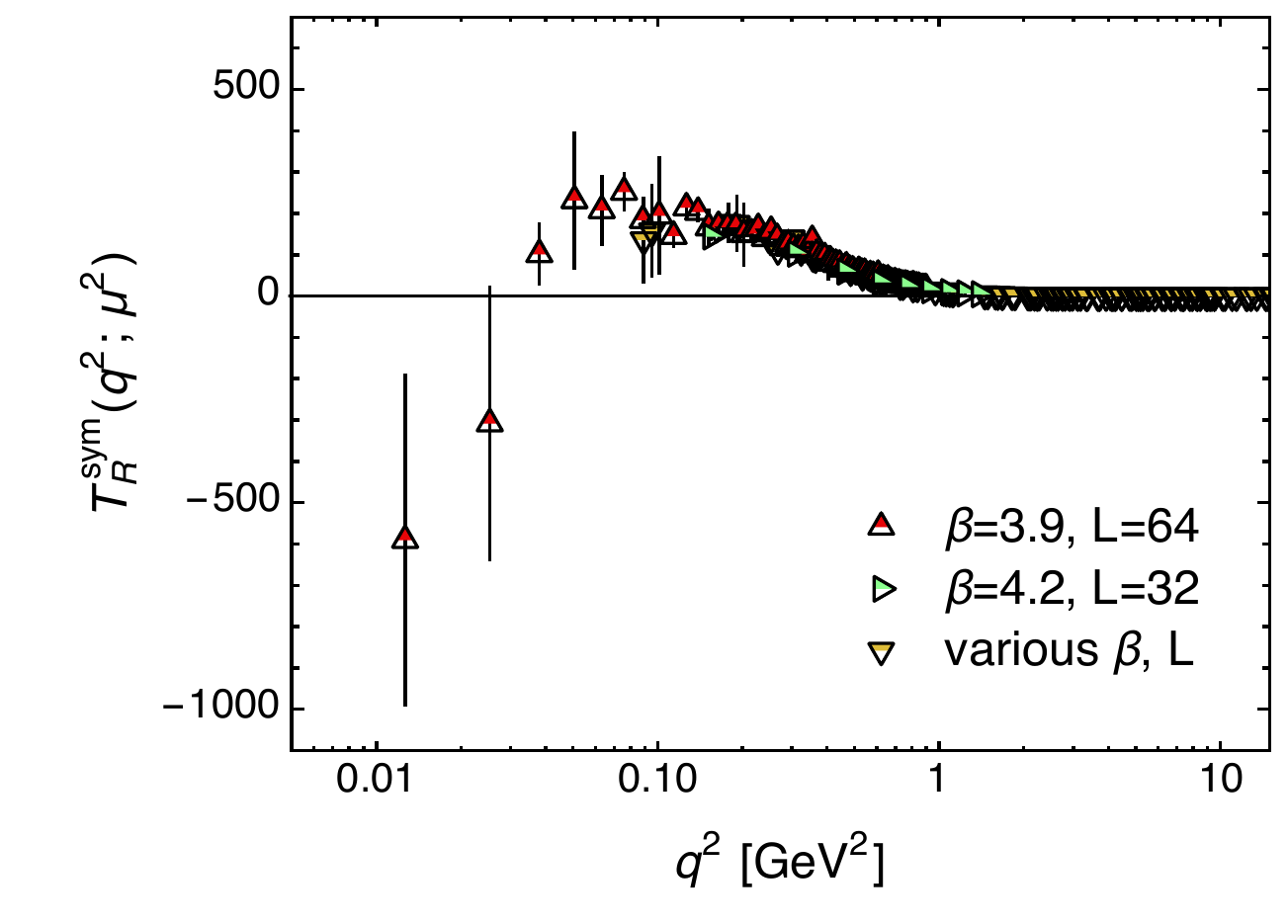}
	\includegraphics[width=0.46\linewidth]{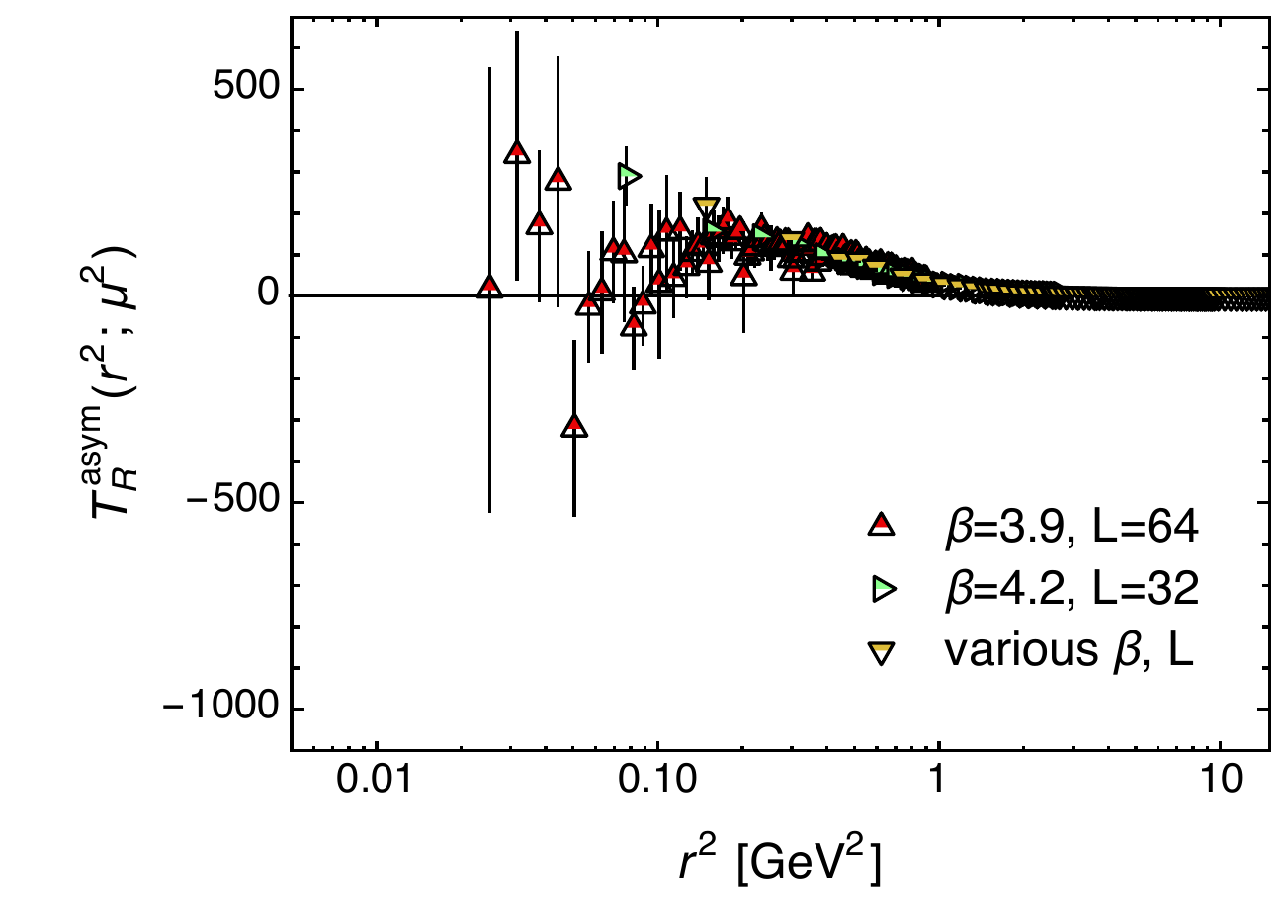}
\end{center}
	\caption{\label{latticeT}(color online) Lattice results for the renormalized connected form factor $\ffT_R$ in the symmetric (left) and asymmetric (right) momentum configuration. For both data sets the renormalization point $\mu=4.3$ GeV was chosen. The same scale is used in both plots which reveals the similar behavior of the two form factors.}
\end{figure*}

In what follows we will consider two special momenta configurations. The first one is the so-called symmetric configuration, in which $q^2=r^2=p^2$ and $q\spr r=q\spr p=r\spr p=-q^2/2$; in this case, there are only two totally transverse tensors, namely
\begin{align}
	\t^{\mathrm{tree}}_{\alpha\mu\nu}(q,r,p)&=\Gamma^{(0)}_{\alpha'\mu'\nu'}(q,r,p)
	 P_{\alpha'\alpha}(q) P_{\mu'\mu}(r) P_{\nu'\nu}(p),\nonumber \\
\t^{S}_{\alpha\mu\nu}(q,r,p)&=(r-p)_{\alpha}(p-q)_{\mu}(q-r)_{\nu}/r^2,
\end{align}  
where $\Gamma^{(0)}_{\alpha\mu\nu}$ is the usual tree-level vertex. Indicating with $\ffSs$ and $\ffTs$ (respectively, $\ffGSs$ and $\ffGTs$) the corresponding form factors in the decomposition of ${\cal G}$ (respectively, $\Gamma$) in this momentum configuration, \1eq{Cto1-PI} implies the relation
\begin{align}
	\ffTs(q^2)&=g\,\Gamma^{\mathrm{sym}}_{T}(q^2)\,\Delta^3(q^2), \nonumber \\
	\ffSs(q^2)&=g\,\Gamma^{\mathrm{sym}}_{S}(q^2)\,\Delta^3(q^2).
\label{Conn-1PI}
\end{align}
In particular, the $\ffTs$ form factor can be projected out through 
\begin{align}
	\ffTs(q^2)&=\left.\frac{W_{\alpha\mu\nu}(q,r,p)\,{\cal G}_{\alpha\mu\nu}(q,r,p)}{W_{\alpha\mu\nu}(q,r,p)W_{\alpha\mu\nu}(q,r,p)}\right\vert_\mathrm{sym},
	\label{prsym}
\end{align}
with $W=\t^{\mathrm{tree}}+\t^{S}/2$.

The second configuration we will study, which will be called `asymmetric' in what follows, is defined by taking the $q\to0$ limit, while imposing at the same time the condition $r^2=p^2=-p\spr r$. In this configuration $\t^{S}_{\alpha\mu\nu}\sim r_\alpha r_\mu r_\nu$ becomes totally longitudinal, and the only transverse tensor one can construct is obtained by the $q\to0$ limit of $\t^{\mathrm{tree}}$ (obviously omitting the $q$ projector), \ie
\begin{align}
	\t^{\mathrm{tree}}_{\alpha\mu\nu}(0,r,-r)&=2r_\alpha P_{\mu\nu}(r).
\end{align}
Thus one is left with a single form factor, which can be projected out through
\begin{align}
	\ffTas(r^2)&=\left.\frac{W_{\alpha\mu\nu}(q,r,p)\,{\cal G}_{\alpha\mu\nu}(q,r,p)}{W_{\alpha\mu\nu}(q,r,p)W_{\alpha\mu\nu}(q,r,p)}\right\vert_\mathrm{asym}\nonumber \\
	&=g\,\ffGTas(r^2)\,\Delta(0)\,\Delta^2(r^2),
	\label{prasym}
\end{align}
where now $W=\t^{\mathrm{tree}}$.

All the quantities defined so far are bare, and a dependence on the regularization cut-off must be implicitly understood. Within a given renormalization procedure, the renormalized Green's functions are calculated in terms of the renormalized fields $\gl_R=\Zgl^{-1/2}\gl$, so that
\begin{align}
	\Delta_R(q^2;\mu^2)&=\Zgl^{-1}(\mu^2)\,\Delta(q^2),\nonumber \\
	\ffTs_R(q^2;\mu^2)&=\Zgl^{-3/2}(\mu^2)\ffTs(q^2),
\end{align} 
and similarly for the asymmetric configuration. Within the MOM scheme that 
we will employ, one then requires that all the Green's functions take their tree-level expression at the subtraction point, 
namely
\begin{align}
	\Delta_R(q^2;q^2)&=\Zgl^{-1}(q^2)\,\Delta(q^2)=1/q^2,\nonumber \\
	\ffTs_R(q^2;q^2)&=\Zgl^{-3/2}(q^2)\,\ffTs(q^2)=g^{\mathrm{sym}}_R(q^2)/q^6.
\end{align}
The first equation yields the renormalization constant $\Zgl$ as a function of the bare propagator, which when substituted 
into the second equation provides a renormalization group invariant definition of the three-gluon MOM running coupling~\cite{Alles:1996ka,Boucaud:1998bq}:
\begin{align}
	g^{\mathrm{sym}}(q^2)&=q^3\frac{\ffTs(q^2)}{[\Delta(q^2)]^{3/2}}=q^3\frac{\ffTs_R(q^2;\mu^2)}{[\Delta_R(q^2;\mu^2)]^{3/2}}.
	\label{alpha3gsym}
\end{align}
In the asymmetric configuration the relation is slightly different, as in this case one has
\begin{align}
	\ffTas_R(r^2;r^2)&=\Zgl^{-3/2}(r^2)\,\ffTas(r^2)=\Delta_R(0;q^2)\,g^{\mathrm{asym}}_R(r^2)/r^4,
\end{align}
implying 
\begin{align}
	g^{\mathrm{asym}}(r^2)&=r^3\frac{\ffTas(r^2)}{[\Delta(r^2)]^{1/2}\Delta(0)}=r^3\frac{\ffTas_R(r^2;\mu^2)}{[\Delta_R(r^2;\mu^2)]^{1/2}\Delta_R(0;\mu^2)}.
	\label{alpha3gasym}
\end{align}
Finally, in both cases the above equations yield for the 1-PI form factors the relation
\begin{align}
	g^i(\mu^2)\,\ffGTR^i(\ell^2;\mu^2)&=\frac{g^i_R(\ell^2)}{[\ell^2\Delta(\ell^2;\mu^2)]^{3/2}},
\end{align}
where $i$ indicates either the symmetric or the asymmetric momentum configuration, and, correspondingly, $\ell^2=q^2,\,r^2$. 

This latter result is of special interest because it establishes a connection between the three-gluon MOM running coupling, which many lattice and continuum studies have paid attention to, and the vertex function of the amputated three-gluon Green's function, a fundamental ingredient within the tower of (truncated) SDEs addressing non-perturbative QCD phenomena. In fact, these quantities are related only by the gluon propagator $\Delta$, which, after the intensive studies of the past decade, is very well understood and accurately known. 

\smallskip

\noindent\textbf{3.$\;$Lattice set-up and results}. The lattice set-up used for our simulations is that of~\cite{Athenodorou:2016gsa}, where quenched SU(3) configurations at several large volumes and different bare couplings $\beta$ were obtained employing the tree-level Symanzik gauge action. In particular, we use 220 configurations at $\beta = 4.20$ for a hypercubic lattice of length $L=32$  (corresponding to a physical volume of 4.5$^4$ fm$^4$) and 900 configurations at $\beta=3.90$ for a $L=64$ lattice (physical volume 15.6$^4$ fm$^4$). The data extracted from these new gauge configurations have been supplemented 
with the one derived from the old configurations of~\cite{Boucaud:2002fx}, obtained using the Wilson gauge action at several $\beta$ (ranging from 5.6 to 6.0), 
lattices (from $L=24$ to $L=32$) and physical volumes (from 2.4$^4$ to 5.9$^4$ fm$^4$).

In~\fig{latticeT} we plot the form factor $\ffT$ renormalized at $\mu=4.3$ GeV for both the symmetric (left panel) and asymmetric (right panel) momentum configuration. In the symmetric case $\ffTs_R$ displays a zero crossing located in the IR region around 0.1--0.2 GeV, after which the data seems to indicate that some sort of divergent behavior manifests itself. In the asymmetric case the situation looks less clear as data are noisier, as a result of forcing one momentum to vanish. 

\smallskip

\noindent\textbf{4.$\;$SDE analysis}. 

In~\cite{Aguilar:2013vaa} it was shown that the {\it nonperturbative} ghost loop diagram contributing to the SDE of $\Delta(q^2)$  
is the source of certain noteworthy effects, the underlying origin of which is the masslessness of the propagators circulating in 
this particular loop.
Specifically, employing a nonperturbative Ansatz for the gluon-ghost vertex that satisfies the correct STI, 
the leading IR contribution, denoted by $\Pi_c(q^2)$, is given by~\cite{Aguilar:2013vaa}   
\begin{align}
\Pi_c(q^2) = \frac{g^2 C_A}{6} q^2 F(q^2) \int_{k}\frac{F(k^2)}{k^2 (k+q)^2} \,,
\end{align}
where  $C_A$ is the Casimir eigenvalue in the adjoint representation, and  $\int_{k}\equiv {\mu^{\epsilon}}/{(2\pi)^{d}}\!\int\!\mathrm{d}^d k$
is the dimensional regularization measure, with $d=4-\epsilon$ and $\mu$ is the 't Hooft mass; evidently, in the limit 
$q^2\to 0$, the above expressions behave like $q^2 \log {q^2}/{\mu^2}$.  Even though this particular term does not interfere with  
the finiteness of  $\Delta(q^2)$, its presence induces two main effects: \n{i}  $\Delta(q^2)$ displays a mild maximum at some relatively 
low value of $q^2$, and  \n{ii} the first derivative of $\Delta^{-1}(q^2)$ diverges logarithmically at $q^2=0$. 
The form of the renormalized gluon propagator that emerges from the complete SDE analysis may be accurately parametrized in the IR by the expression  
\begin{align}
	\Delta_R^{-1}(q^2;\mu^2)\underset{q^2\to0}{=}q^2\left[a+b\log\frac{q^2+m^2}{\mu^2}+c\log \frac{q^2}{\mu^2} 
	\right]+m^2,
	\label{modindDelta}
\end{align}   
with $a$, $b$, $c$, and $m^2$ suitable parameters, which captures explicitly the two aforementioned effects. Note that  
\mbox{$\Delta_R^{-1}(0;\mu^2) =m^2$}, and that the `protected' logarithms stem from gluonic loops.  

Higher order $n$-point functions ($n>2$) are also affected~in~notable ways by the presence of ghost loops in their
diagrammatic expansion\footnote{We refer to ghost loops that exist already at the one-loop level. Ghost loops nested within gluon loops do not produce particular effects, because the additional 
integrations over virtual momenta soften the IR divergence.}. 
If the external legs correspond to background gluons (as was the case in~\cite{Aguilar:2013vaa}), the leading 
IR behavior of projectors such as~\noeq{prsym} and/or~\noeq{prasym} is proportional to the derivative 
of the inverse gluon propagator~\cite{Aguilar:2013vaa}, by virtue of the Abelian STIs.
Thus, eventually, a logarithmic divergence appears, which drives the aforementioned projectors 
from positive to (infinitely) negative values, causing invariably the appearance of a zero crossing. 
Use of the `background quantum' identities~\cite{Grassi:1999tp,Binosi:2002ez}, which relate 
background Green's functions with quantum ones, reveals that the same behavior is expected for quantum external legs, modulo a (finite) 
function determined by the ghost-gluon dynamics~\cite{Aguilar:2013vaa}. 
The exact position of the zero crossing is 
difficult to estimate, because it depends on the details of all finite contributions that are `competing' against the 
logarithm coming from the ghost loop; however, it is clear that the tendency, in general, is to appear in the deep IR.  

\begin{figure}
	\hspace{-0.5cm}\includegraphics[width=0.96\linewidth]{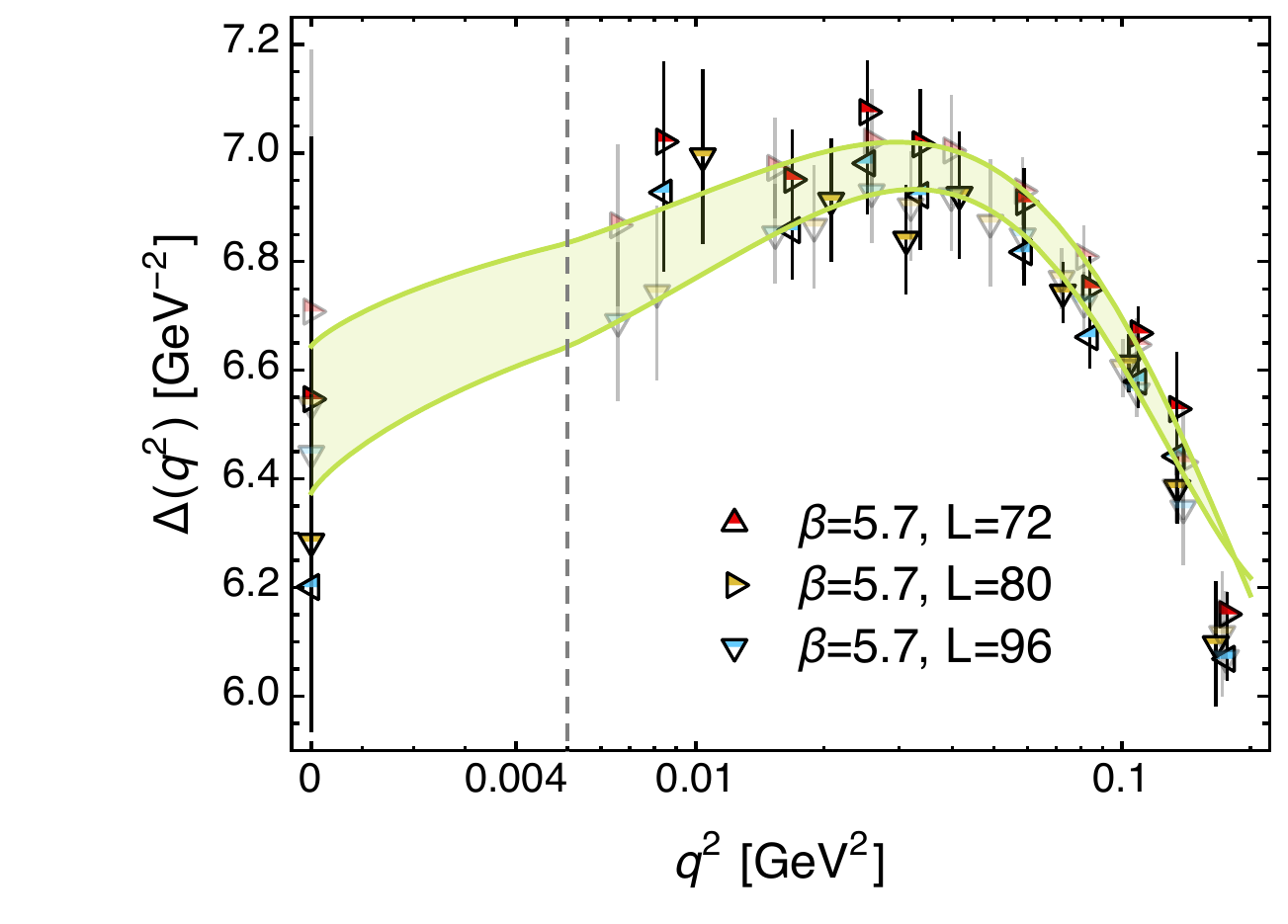}
	\caption{\label{IRDelta}(color online) IR fit of the gluon propagator at $\mu=4.3$ GeV. The actual data fitted, are the semitransparent ones, which are obtained by applying a cubic smooth spline with roughness penalty procedure to the original data~\cite{Green:1994}. The band indicates the variation of the fit between $L=72$ and $L$~$=96$, $L=80$ being somewhere in between. Notice also that after the dashed vertical line the scale becomes linear, to expose the propagator behavior at the origin.}\vspace{-.1cm}
\end{figure}

We emphasize that independent analyses within the SDE formalism, employing a variety of techniques and truncation schemes,
have confirmed these claims in the three-point~\cite{Blum:2014gna,Eichmann:2014xya,Cyrol:2016tym} and the four-point~\cite{Binosi:2014kka,Cyrol:2014kca} gluon sector. In addition, unquenching techniques, such as those developed in~\cite{Aguilar:2012rz,Aguilar:2013hoa}, coupled with the lattice results of~\cite{Ayala:2012pb}, show that the presence of light quarks slightly modifies the behavior of the gluon and ghost two-point sector only at the quantitative level; therefore, the expected IR pattern `zero crossing plus logarithmic IR divergence' for $n$-point gluon Green's functions seems  
to constitute a robust prediction {\it for QCD}\footnote{Note, however, that at least in the case of the three-gluon vertex, a preliminary study~\cite{poster} shows that light quarks will move the zero crossing deeper in the IR, which might render it undetectable in current full QCD lattice simulations.}. 

In particular, for the form factors under scrutiny, one expects the (configuration independent) IR behavior
\begin{align}
	\ffGTR^i(\ell^2;\mu^2)&\underset{\ell^2\to0}{\simeq}F(0;\mu^2)\frac{\partial}{\partial\ell^2}\Delta_R^{-1}(\ell^2;\mu^2),
	\label{IRGammaT}
\end{align}
where $F(0)\approx2.9$ at $\mu=4.3$ GeV~\cite{Bogolubsky:2009dc}.

To see if indeed the lattice data conform to the expected behavior, we start by estimating the propagator's parameters $a$, $b$, $c$ and $m^2$ by fitting the lattice data of~\cite{Bogolubsky:2009dc}. The results are shown in \fig{IRDelta}, with the parameter values obtained for the available data sets listed in Table~\ref{bestfitvalues}. In what follows we will not distinguish between these different fits; rather we will use a single curve with bands representing its `uncertainty'. 

\begin{table}[!t]
\begin{center}
\begin{tabular}{cccc}
\hline\hline
{Parameter}&\hspace{.85cm} {$L=72$}\hspace{.85cm} & $L=80$& \hspace{.85cm} $L=96$  \\
\hline
$a$ & -0.471 & -0.151 &\hspace{.85cm} -1.146 \\
$b$ & -0.546 & -0.458 &\hspace{.85cm} -0.922 \\
$c$ & 0.362 & 0.352 &\hspace{.85cm} 0.546\\
$m^2$ & 0.151 & 0.154 &\hspace{.85cm} 0.157 \\
\hline
\end{tabular}
\end{center}
\caption{\label{bestfitvalues} Best fit parameters for the IR propagator~\noeq{modindDelta} obtained using the SU(3) data of~\cite{Bogolubsky:2009dc} for $\beta=5.7$ and $L=72$, $80$ and $96$ lattices.}
\end{table}

\begin{figure*}[!t]
\begin{center}
\hspace{-.75cm}
	\includegraphics[width=0.46\linewidth]{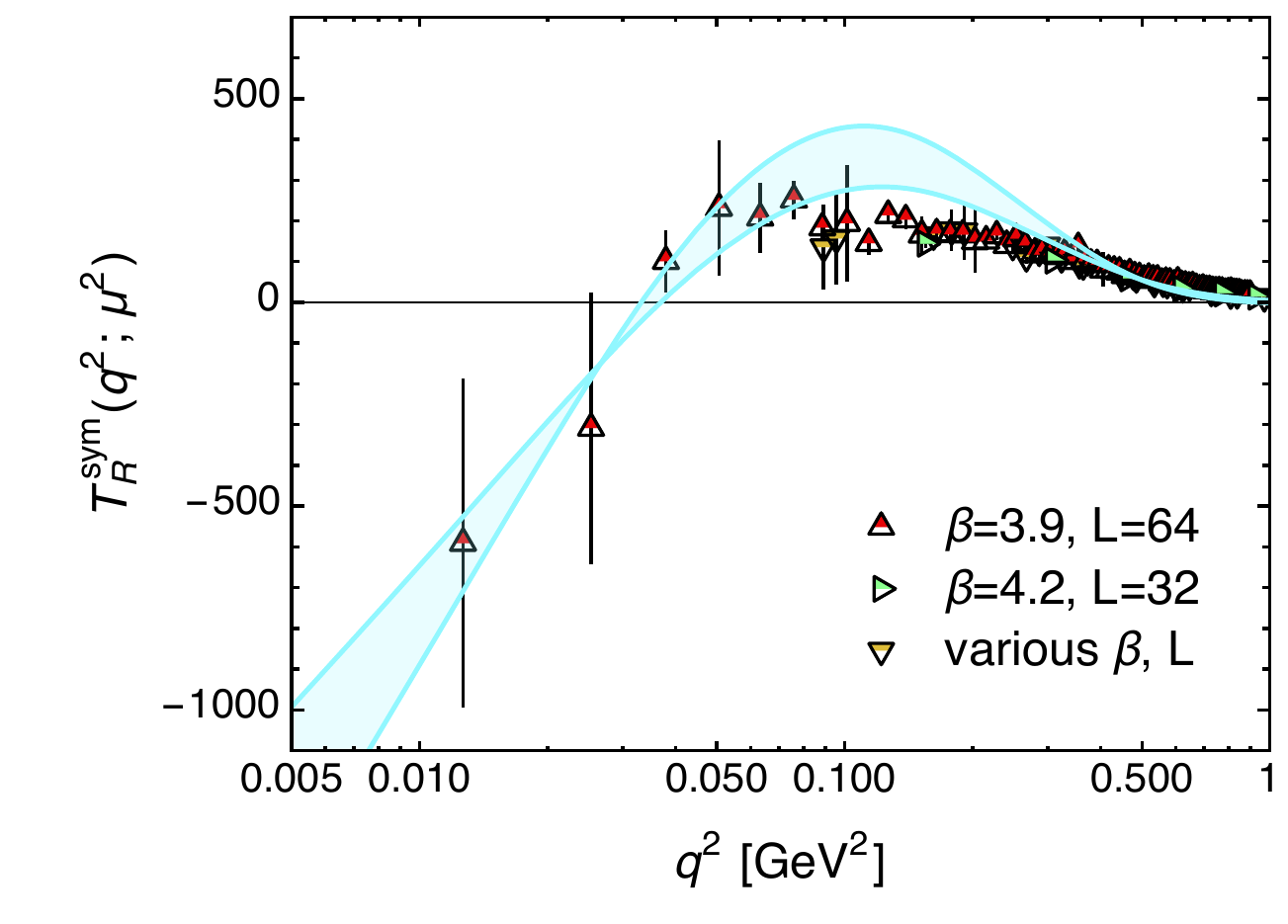}
	\includegraphics[width=0.46\linewidth]{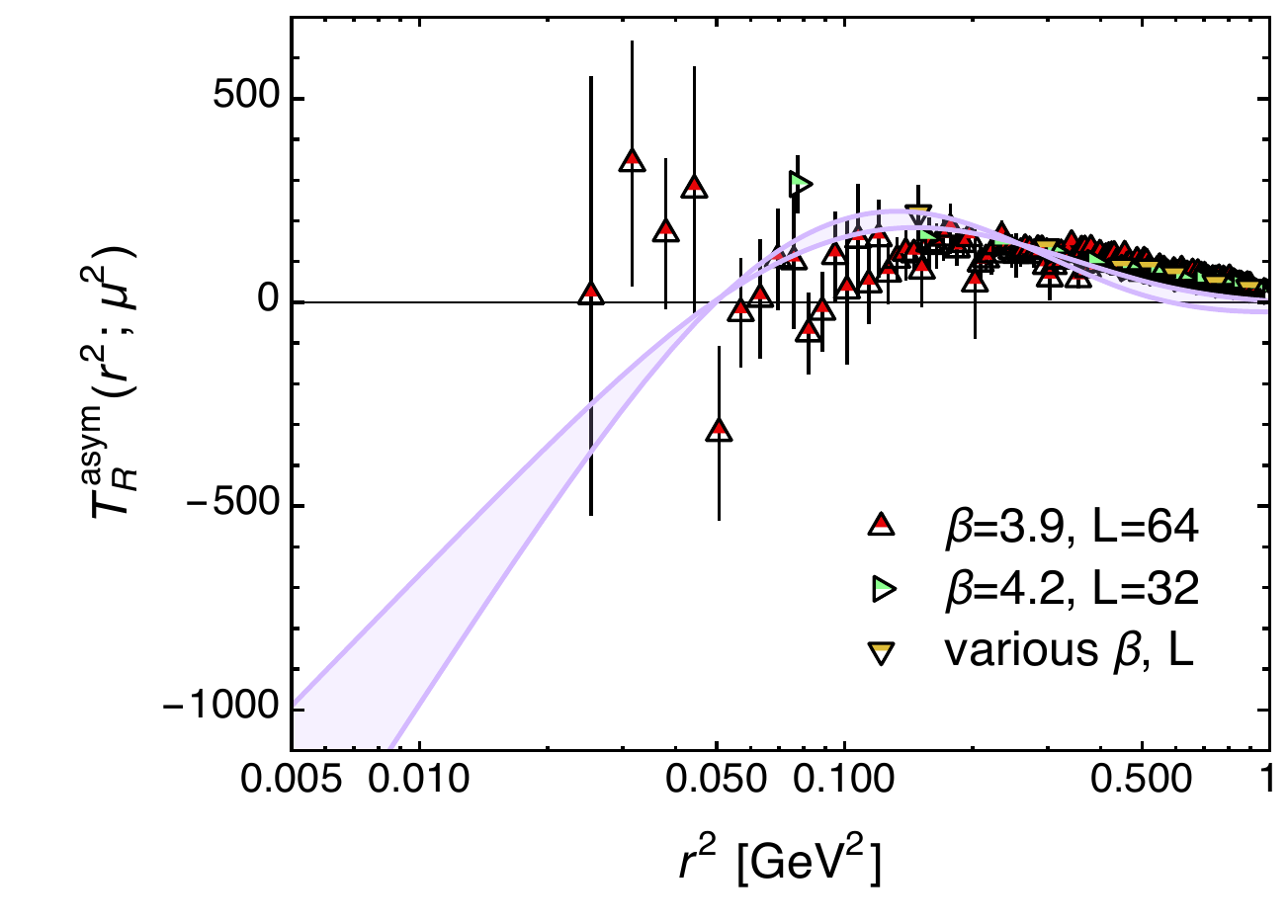}
	\end{center}
	\begin{center}
	\hspace{-.75cm}
	\includegraphics[width=0.46\linewidth]{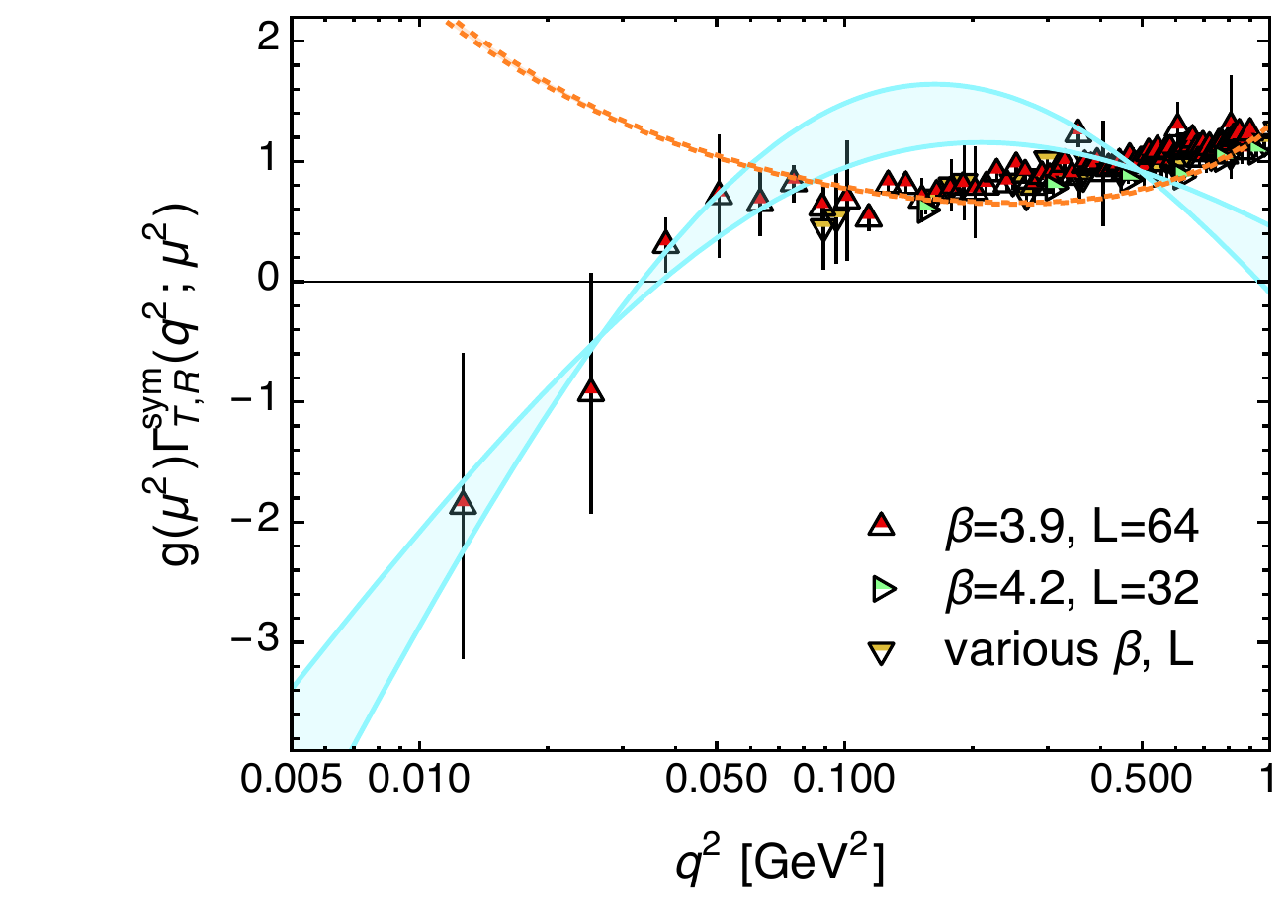}
	\includegraphics[width=0.46\linewidth]{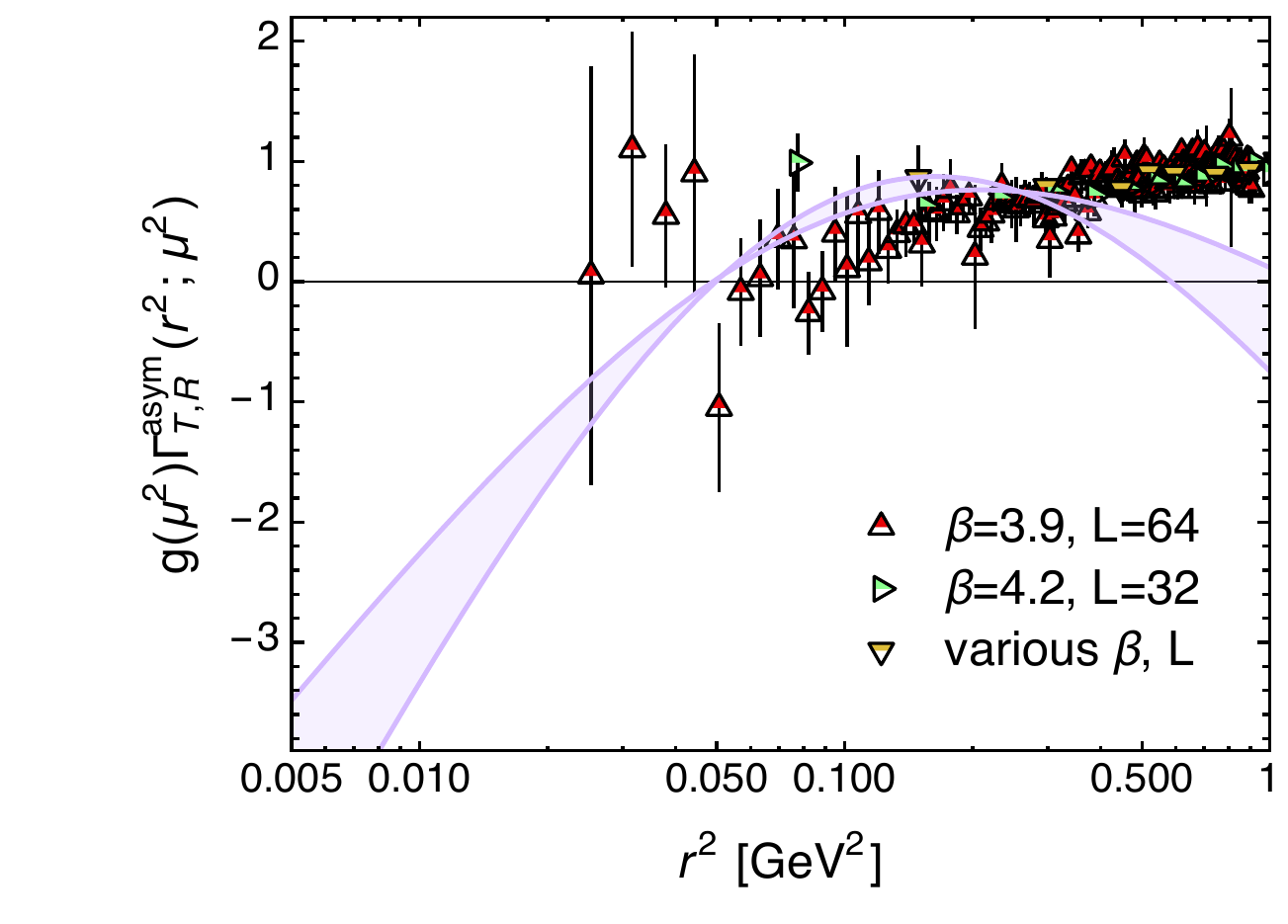}	
	\end{center}
	\caption{\label{compT}(color online) Comparison between the lattice results for the renormalized connected and 1-PI form factors $\ffT_R$ and $g\ffGTR$ and the SDE prediction in the symmetric (left panels) and asymmetric (right panels) configurations. The band, as in~\fig{IRDelta}, appears bounded by the results obtained with the fits of the lattice propagators for $L=72$ and $L=96$, and aims at giving an indication of the variation of the results. To ease the comparison, left and right panels have the same scale. Zero crossing happens at $190$ and $220$ MeV respectively. For the quantity $\ffGTR^{\mathrm{sym}}$ (lower left panel) we also plot (dashed line) the semiclassical approximation~\noeq{instantons}.}
\end{figure*}

At this point we can use~\1eq{IRGammaT} and the relation~\noeq{Conn-1PI} to determine the expected IR behavior of the connected form factors $\ffTs$ and $\ffTas$, and compare with the data\footnote{Subleading terms are collectively taken into account by adding an extra constant in~\noeq{IRGammaT} the value of which is then determined by refitting the data (only for this parameter).}. The results are shown in~\fig{compT}. While it is evident that in the symmetric case a good description of the IR data is achieved, in the asymmetric case the positive excess in the data coupled to the large errors make it more difficult to discern the low momentum behavior of $\ffTas_R$ and $\ffGTR^{\mathrm{asym}}$. 

There is an interesting conclusion one might draw from the behavior of these form factors. As discussed in detail in~\cite{Athenodorou:2016gsa}, when quantum fluctuations can be either neglected or suppressed, gluon correlation functions appear to be dominated by a semiclassical background described in terms of a multi-instanton solution. In particular, specializing to the symmetric configuration, one has in this case
\begin{align}
	g^{\rm{sym}}(\mu^2) \ffGTR^{\mathrm{sym}}(q^2;\mu^2)\simeq \sqrt{
	\frac{2}{9 n p^2 \left[ \Delta(p^2;\mu^2)\right]^3}},
	\label{instantons}
\end{align}
where $n=7.7$ fm$^{-4}$ is the instanton density in the semiclassical background. The resulting curve is shown by the dashed line in the lower left panel of~\fig{compT}. Then, we see that while the approximation~\noeq{instantons} appears to be justified for momenta roughly below $q\sim1$ GeV~\cite{Boucaud:2002fx,Athenodorou:2016gsa}, it fails in the deep IR region, around $q\sim0.2$--$0.3$ GeV. This can be understood once we notice that at such low momenta (where the zero crossing takes place), the dynamics is entirely dominated by  massless ghosts; plainly, this is a quantum effect that cannot be captured within the framework of a semiclassical approach. 

\begin{figure*}[!t]
\begin{center}
\hspace{-0.75cm}
	\includegraphics[width=0.46\linewidth]{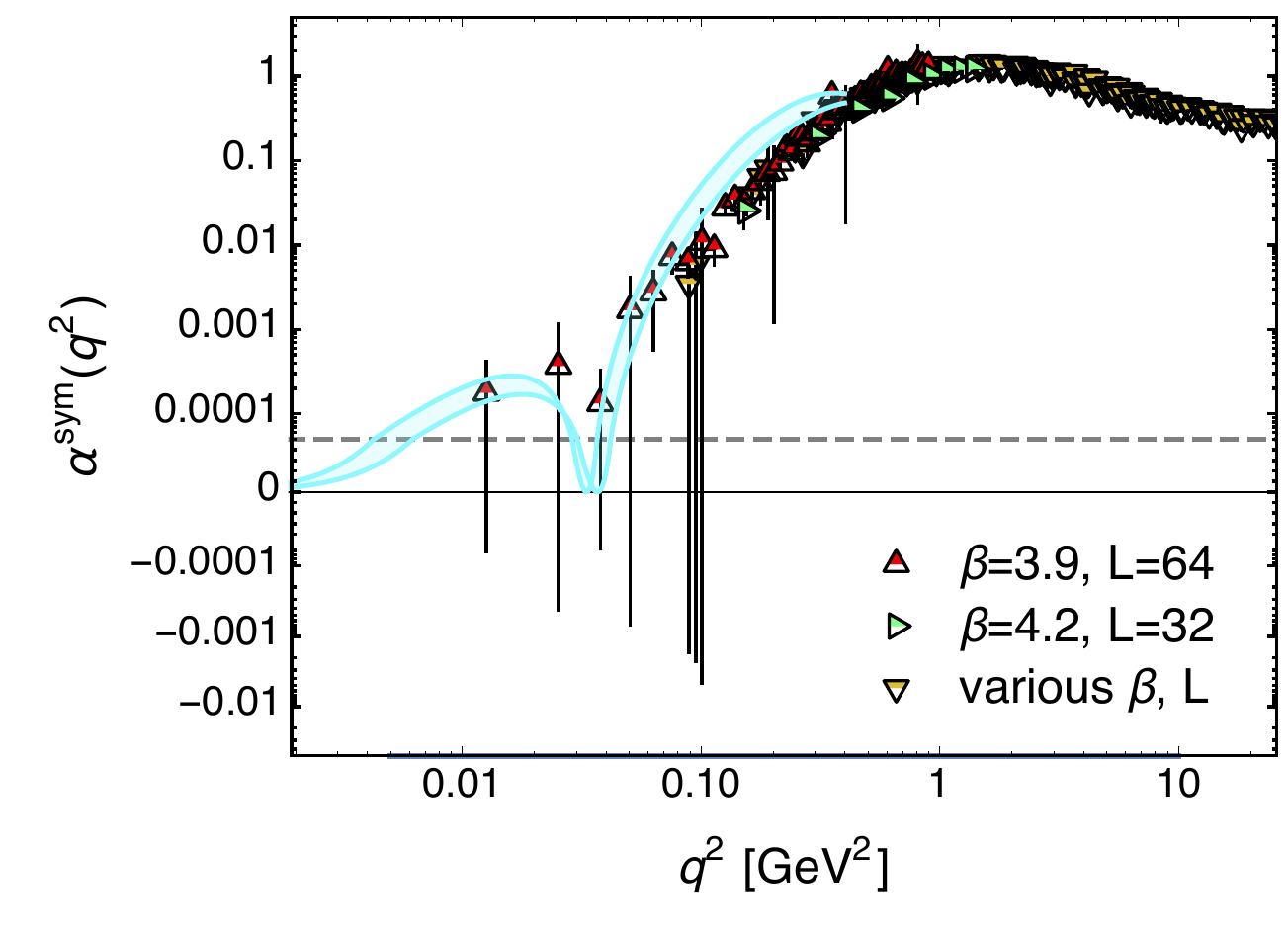}
	\includegraphics[width=0.46\linewidth]{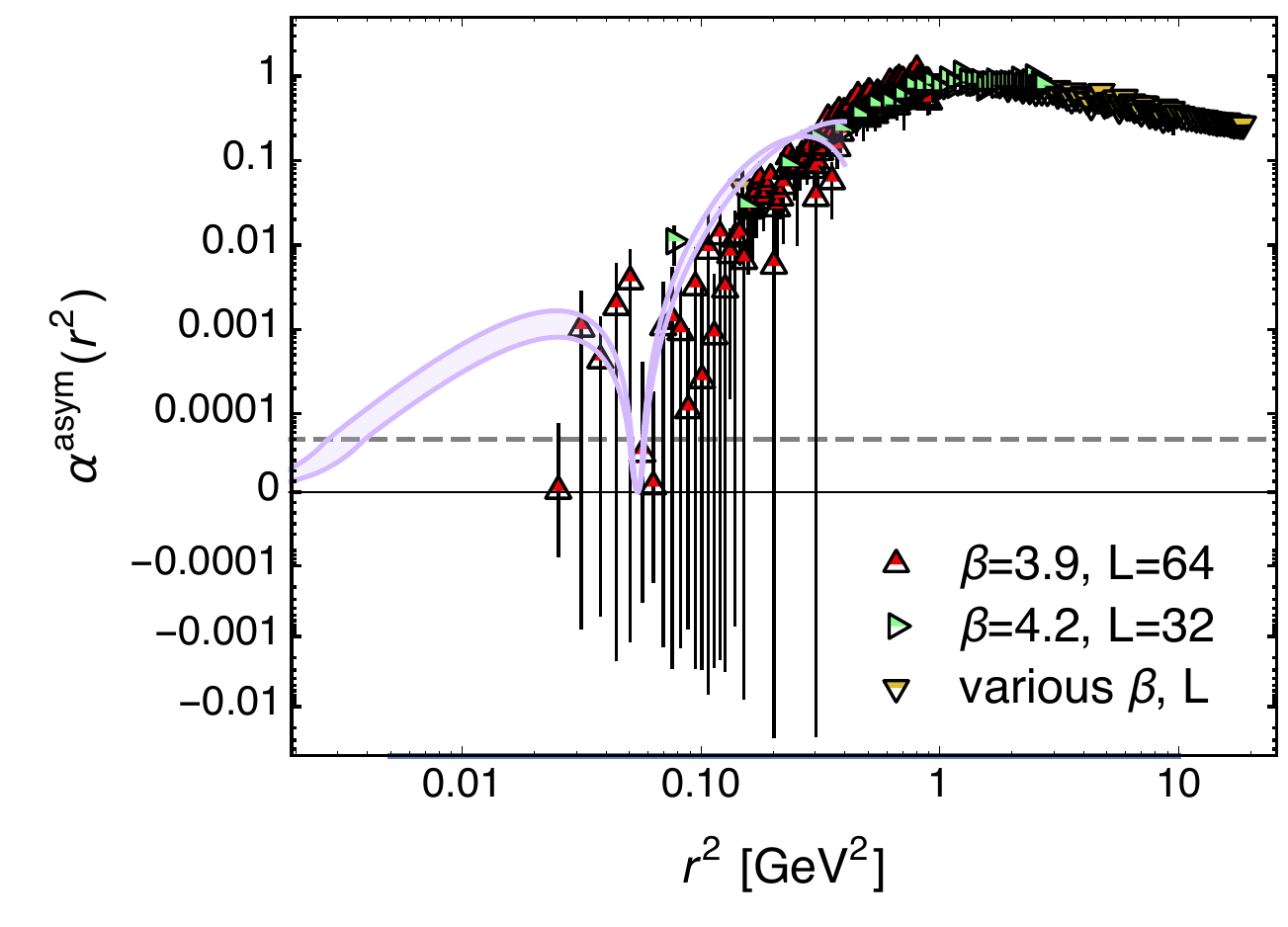}
\end{center}	
	\caption{\label{alpha3g}(color online) Comparison between the lattice results for the three gluon effective coupling and the SDE prediction in the symmetric (left) and asymmetric (right) configuration. Notice that on the $y$ axis scale switch from logarithmic to linear at the location of the dashed gray line (and then back to logarithmic for $y<0$); while this choice exaggerates the error bars, it has the advantage of exposing the vanishing of the coupling at a non vanishing momentum~value.}
\end{figure*}

Next, using \2eqs{alpha3gsym}{alpha3gasym}, we can construct the effective coupling $\alpha^i(\ell^2)=g^{i2}(\ell^2)/4\pi$ both from the lattice data and the determined IR behavior. 
In particular, the $\alpha^i$ derived from the three-gluon vertex is proportional 
to the {\it square} of the form factor $\ffGT$, and displays a striking behavior: 
$\alpha^i$ is forced to vanish at the zero crossing,  and then `bounces' back to positive values, as 
can be clearly seen in~\fig{alpha3g}. According to this result, the part of the amplitude  
`gluon + gluon $\to$ gluon + gluon' that is mediated 
by the (fully dressed) one-gluon exchange diagram vanishes at some special IR momentum; to be sure, this is not true for the 
entire physical amplitude, since additional diagrams (such as `boxes') will furnish nonvanishing contributions.

\smallskip

\noindent\textbf{4.$\;$Conclusions.} We have presented new lattice results for the three-gluon vertex form factor $\ffT$ proportional to the tree-level tensor structure. The data were obtained from large 4-dimensional volumes configurations generated for an SU(3) Yang-Mills theory gauge fixed in the Landau gauge, and the form factor evaluated in the so-called symmetric and asymmetric momentum configurations. The IR behavior of $T$ was then scrutinized in detail and contrasted with (model independent) SDE predictions finding good agreement. In doing so, we have discussed, the failure of a semiclassical picture based on instantons due to the quantum effects associated to massless ghost loops.

It is tempting to speculate that the behavior seen in the deep IR in the asymmetric case is not entirely due to statistical fluctuations. In fact, it has been shown in~\cite{Aguilar:2016vin} that in this momentum configuration the 1-PI form factor receives contributions from quantities describing the appearance of (longitudinally coupled) massless poles in the fundamental vertices of the theory. When such poles are present, in fact, the (Abelian) STIs acquires new terms that account for both the IR finiteness of the gluon propagator and the deformation of~\1eq{IRGammaT}, which would now read~\cite{Aguilar:2016vin}
\begin{align}
	\ffGTR^{\mathrm{asym}}(\ell^2;\mu^2)&\underset{\ell^2\to0}{\simeq}F(0;\mu^2)\left[\frac{\partial}{\partial\ell^2}\Delta_R^{-1}(\ell^2,\mu^2)-C'_1(r^2;\mu^2)\right],
	\label{IRGammaTpoles}
\end{align}
where $C'_1$ corresponds, modulo a numerical factor, to the wave function amplitude associated to the formation of the massless pole. The shape of $C'_1$ and the corresponding qualitative modification induced to the form factor has been sketched in~\cite{Aguilar:2016vin}; the expected signal is a positive excess in the IR region which is very similar to what the data in~\fig{compT} show. 

\bigskip 

\noindent{{\it Acknowledgements.}} The research of J.P. and J. R-Q is supported by the Spanish MINECO under grant 
FPA2014-53631-C2-1-P and FPA2014-53631- C2-2-P and SEV-2014-0398, and Generalitat Valenciana under grant Prometeo II/2014/066. S. Z. acknowledges support by the Alexander von Humboldt foundation. We thank K. Cichy, M. Creutz, O. P\`ene, O. Philipsen, M. Teper, J. Verbaarschot for fruitful discussions. Numerical computations have used resources of CINES and GENCI-IDRIS as well as resources at the IN2P3 computing facility in Lyon. 


\end{document}